\input{aipcheck}

\documentclass[
    ,final            
  ]
  {aipproc}

\layoutstyle{6x9}

\begin{document}

\title{Rapid GRB Follow-up with the 2-m Robotic Liverpool Telescope}


\author{Andreja Gomboc}{
  address={Astrophysics Research Institute, Liverpool John Moores University, UK}
  ,altaddress={University in Ljubljana, Slovenia}
}

\author{Michael F. Bode}{
  address={Astrophysics Research Institute, Liverpool John Moores University, UK}
}

\author{David Carter}{
  address={Astrophysics Research Institute, Liverpool John Moores University, UK}
}
\author{Cristiano Guidorzi}{
  address={Astrophysics Research Institute, Liverpool John Moores University, UK}
}
\author{Alessandro Monfardini}{
  address={Astrophysics Research Institute, Liverpool John Moores University, UK}
}
\author{Carole G. Mundell}{
  address={Astrophysics Research Institute, Liverpool John Moores University, UK}
}
\author{Andrew M. Newsam}{
  address={Astrophysics Research Institute, Liverpool John Moores University, UK}
}
\author{Robert J. Smith}{
  address={Astrophysics Research Institute, Liverpool John Moores University, UK}
}
\author{Iain A. Steele}{
  address={Astrophysics Research Institute, Liverpool John Moores University, UK}
}
\author{John Meaburn}{
  address={University of Manchester, UK}
}

\begin{abstract}
 We present the capabilities of the 2-m robotic Liverpool Telescope (LT), owned and operated by 
 Liverpool John Moores University and situated at ORM, La Palma.
 Robotic control and scheduling of the LT make it especially powerful for observations in 
 time domain astrophysics including: (i) rapid response to Targets of Opportunity: Gamma Ray Bursts, novae,  supernovae, 
 comets; (ii) monitoring of variable objects on timescales from seconds to years, and (iii) 
 observations simultaneous or coordinated with other facilities, both ground-based and from space.
 Following a GRB alert from the Gamma Ray Observatories HETE-2, INTEGRAL and Swift 
 we implement a special over-ride mode which enables observations 
 to commence in about a minute after the alert, including optical and near infrared imaging and 
 spectroscopy. In particular, the combination of aperture, site, instrumentation and rapid response 
 (aided by its rapid slew and fully-opening enclosure) makes the LT excellently suited to help 
 solving the mystery of the origin of optically dark GRBs, for the investigation of short bursts 
 (which currently do not have any confirmed optical counterparts) and for early optical spectroscopy 
 of the GRB phenomenon in general. 
We briefly describe the LT's key position in the RoboNet-1.0 network of robotic telescopes.
\end{abstract}

\maketitle


\section{Early GRB Afterglows in Optical and Infra-Red}
\subsection{Bright Optical Afterglows of Long GRBs}
 GRB990123 is the famous case of a prompt optical flash. Although detected more than 5 years ago it is still 
one of the very few (so far, only 4) optical afterglows (GRB 990123, 021004, 021211, 030418) 
that were detected 
a few minutes after the gamma ray burst. 
However, early GRB afterglows are essential in the study of extreme physics
of ultra-relativistic flows and shocks and provide unique probes of the circum-burst 
medium and the nature of GRB progenitors.

In general, optical afterglows fade according to a power law $F\sim t^{\alpha}$
with power law index $\alpha$ between -0.6 and -2.3 \cite{Grei04}.
But in some cases, such as GRB 021004, dense photometric coverage revealed
additional light curve structure on short timescales as well as colour changes 
in early optical afterglows. 
According to different models such fluctuations could be related to phenomena in jets, circum-burst medium
or renewed activity of the central engine \cite{Bers03}, \cite{Lazz02}. 

Among the most important issues regarding the origin and overall energetics of GRBs are achromatic breaks
observed in some light curves a few hours to a few days after the GRB. Interpreted as due to beaming, 
they indicate jet opening angles of a few degrees and have led to evidence that most GRBs have a standard energy
reservoir \cite{Frai01}.
In a number of cases though, the exact time of the break is controversial due to insufficiently dense
sampling of the light curve.

\subsection{Optically Dark GRBs}
Contrary to X-ray afterglows, which are detected in most well localized GRBs, 
optical afterglows were so far observed in approximately half of them. These missing afterglows are 
usually referred to as "optically dark".
Although small robotic telescopes provide fast magnitude upper limits, these usually do not go deep enough
to be conclusive. It is still an open question whether 
dark optical afterglows are merely observationally overlooked or not detected due to some of their properties.
Their non-detection could be, for example, due to the fact that they are 
intrinsically faint or initially bright but very rapidly fading. 
Other explanations include heavy dust obscuration in host galaxies or
their position at high redshifts of z=5 - 10.
The latter possibility is extremely interesting since GRBs at such high redshifts would provide a unique 
cosmological probe of star formation and galaxy evolution in the early universe. 
To solve the mystery of dark bursts,  
rapid observations in infrared wavelengths are of key importance.

\subsection{Afterglows of Short GRBs}
For short GRBs no unambiguous afterglow has been
detected so far and their nature remains an enigma. 
The favourite model is a binary neutron star - neutron star or
neutron star - black hole merger, although no evidence is available to prove or disprove this. 
It has been predicted \cite{Pana01} 
that their early optical afterglows may be much fainter than those of long GRBs:
20 - 23 magnitude initially and fading to 25 - 28 magnitude after a day, indicating that rapid and
deep follow-up in the optical may again play a crucial role.

\subsection{X-ray Flashes}
A special class of objects, which seem to be related to GRBs are X-Ray Flashes (XRFs): they are similar
to GRBs in many characteristics of their prompt emission with the main difference that they peak in X-rays
instead of gamma rays. 
In the optical, several afterglows (XRF 020903, XRF030723 \cite{Sode02}, \cite{Fox03}) 
and host galaxies (XRF 011030, XRF 020427 \cite{Bloo03}) were detected. 
More observations in the near future will hopefully contribute to clarifying the XRF and GRB 
similarities and relationship and, 
on the other hand, also the issue of their differences: i. e. whether they differ due to different total energy, 
structure of the jet, progenitor's size, redshift, etc.

\vspace{0.5cm}
In view of the many open issues briefly outlined here and the fact that only a handful of optical 
afterglows were observed within a 
few minutes or even an hour after the GRB, it is obvious that early 
multi-colour optical and infrared photometry and spectroscopy can provide valuable information and help 
better to understand the nature of GRBs and their afterglows.

\section{The Liverpool Telescope}
\begin{figure}
  \includegraphics[height=.29\textheight]{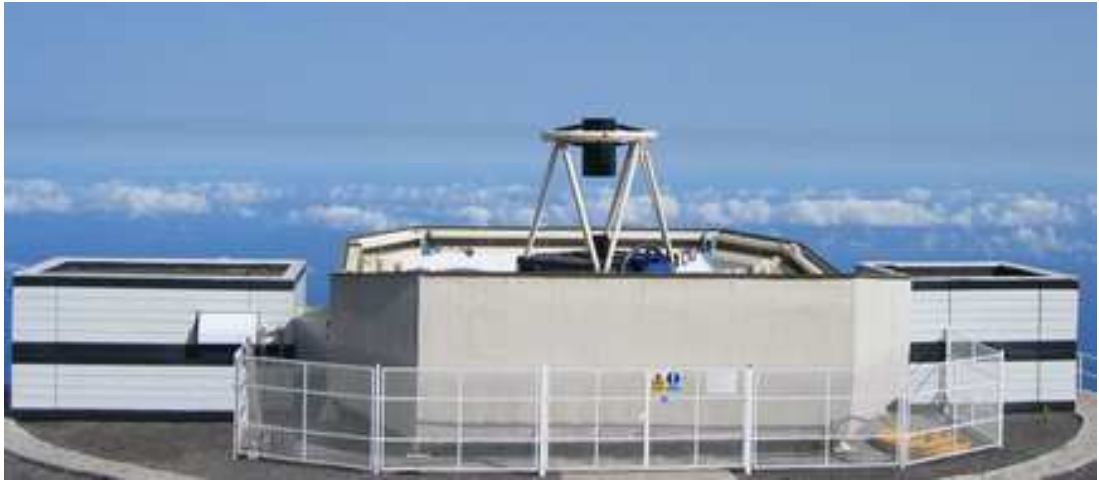}
      \caption{The Liverpool Telescope at ORM, La Palma, Canary Islands.}
      \label{LT}
\end{figure}

The Liverpool Telescope (LT, see Fig. \ref{LT}) is situated at Roque de los Muchachos on La Palma in the Canaries and is operated by Liverpool 
John Moores University as a National facility under the auspice of PPARC. One of its noticeable characteristics is the clam-shell enclosure, which is fully opening. 
The benefits of such an enclosure are the minimization of the dome seeing, fast thermal equilibrium (reached in less than 30 min) and,
particularly important in GRB follow-up, the short response time, since there is no need to wait for the dome to slew.
Drawbacks are 
potential for windshake and the fact that in case of the enclosure breakdown, the telescope is totally
exposed to weather conditions. Windshake is minimised by the 'stiff' structure of the telescope design and the enclosure has 
battery back-up in the event of power failure.

Other telescope specifications are: 2-m primary mirror, final focal ratio f/10,
altitude-azimuth design, image quality < 0.4" on axis, pointing < 2arcsec rms,
slew rate of 2$^o$/sec, five instrument ports (4 folded and one straight-through, selected by a deployable, 
rotating mirror in the AG Box within 30s) and robotic (unmanned) operation with automated scheduler.

\begin{table}[htb]
\caption{The Liverpool Telescope instrumentation }
\label{tableLT}
\begin{tabular}{ll}
\hline
{\it RATCam} Optical CCD Camera - & 2048$\times$2048 pixels, 0.135"/pixel, FOV 4.6'$\times$4.6', \\
  & 8 filter selections (u', g', r', i', z', B, V, ND2.0) \\
 & - from LT first light, July 2003 \\
 \hline
 {\it SupIRCam} 1 - 2.5 micron Camera - & 256$\times$256 pixels, 0.4"/pixel, FOV 1.7'$\times$1.7', \\
  (with Imperial College) & Z, J, H, K' filters - from late 2004 \\
  \hline
 {\it Prototype Spectrograph} - & 49, 1.7" fibres, 512$\times$512 pixels, R=1000; \\
  (with University of Manchester) & 3500 $<$ $\lambda$ $<$ 7000 \AA - from 2005 \\
  \hline
 {\it FRODOSpec} Integral field  & R=4000, 8000; \\
  double beam spectrograph - & 4000 $<$ $\lambda$ $<$ 9500 \AA - from 2006 \\
  (with University of Southampton) & \\
\hline
\end{tabular}\\[2pt]
\end{table}

At present, instrumentation (Table \ref{tableLT}) includes Optical and Infrared cameras. A prototype low resolution 
spectrograph will be commissioned in 2005 and a higher resolution spectrograph is being developed for 2006.

The telescope began science operations in January 2004 and after the enclosure hydraulics 
upgrade in summer 2004, the LT is entering fully robotic mode. It is expected that the telescope 
will operate in a fully autonomous way (without human intervention) from the beginning of 2005.

Science programmes running on the LT are diverse, but due to
robotic control \cite{Fras02} and automated scheduling \cite{Stee97}, \cite{Fras04} the LT is 
especially suited for: 
\begin{itemize}
\item{rapid response to Targets of Opportunity (GRBs, novae,  supernovae, comets);}
\item{monitoring of variable objects on timescales from seconds to years 
	(follow-up of the ToO, active galactic nuclei, gravitational lenses etc.);}
\item{observations simultaneous or coordinated with other facilities, both ground-based and from space;}
\item{condition (e.g. seeing, photometricity) dependent or time critical (e.g. binary phase) 
observations.}
\end{itemize}

\begin{figure}
  \includegraphics[height=.44\textheight]{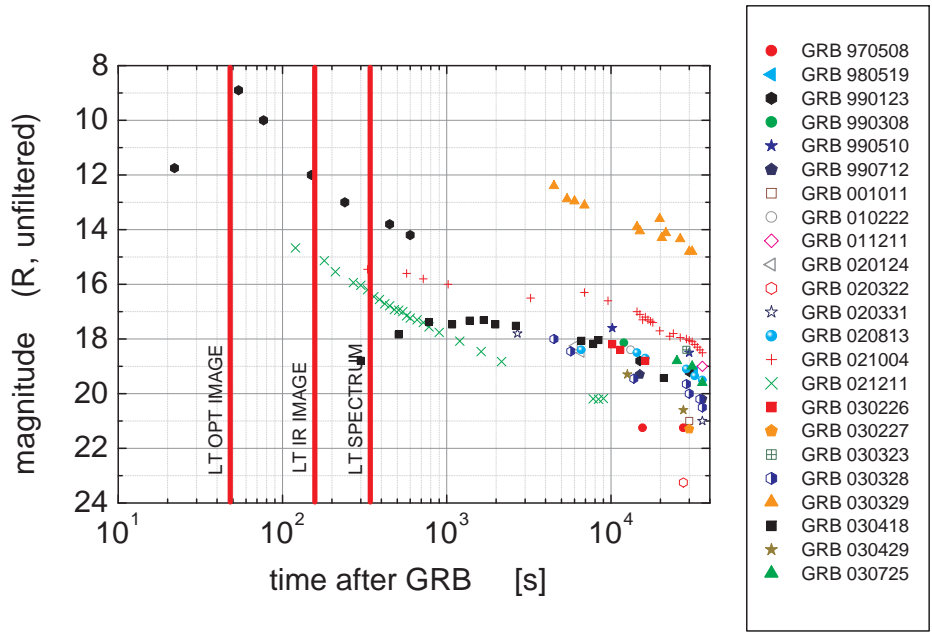}
      \caption{Optical afterglows detected in first minutes and hours after the GRB. 
      Vertical lines indicate typical times of first observations with LT instrumentation.
      (Magnitudes and times are taken from GCN Circulars and are intended for illustrative 
      purpose only.)
}
\end{figure}

\section{Rapid GRB Follow up observations with the Liverpool Telescope}

The GRB programme on the LT takes advantage of the telescope's robotic control:
following receipt of GRB alert from the GCN, the Robotic Control System interrupts ongoing observations,
applies over-ride
mode and starts with GRB observations according to the basic GRB follow-up strategy, 
which presented in simple terms proceeds through following steps:

\begin{itemize}
\item{slew to the position given in GCN alert;}
\item{start optical imaging in about 1 min;}
\item{try to identify the optical transient by comparison with the USNO catalogue or by image subtraction;}
\item{if no candidate afterglow in the optical is detected, continue observations in the infrared;}
\item{if an optical transient is reliably identified:}
\begin{itemize}
   \item{continue with multicolour imaging with intervening spectroscopy (if the optical transient 
   is brighter than magnitude 15 for prototype spectrograph and 19 for
FRODO spectrograph);}
   \item{issue a GCN Circular;}
   \item{trigger larger facilities.}
\end{itemize}

\end{itemize}

The advantages of the LT in comparison with smaller robotic telescopes in GRB follow-up are larger aperture and
deeper observations, the number of filters, infrared
imaging and the possibility of early spectroscopy. These make it particularly suitable for study of: 
\begin{itemize}
\item{afterglows of short GRBs,}
\item{afterglows of optically dark GRBs,}
\item{prompt optical flashes,}
\item{early optical GRB spectrometry, and}
\item{statistical properties of GRBs and their afterglows}.
\end{itemize}

With approximately 25 percent of GRBs occurring at night over La Palma and 70$^\circ$ maximal zenith distance
observable by the LT, we expect to follow-up 1 in 6 GRBs immediately following the alert.
We plan to monitor GRB afterglows at later stages depending on their scientific significance and in 
collaboration with other facilities, including the Faulkes Telescopes as part of RoboNet-1.0.  

\section{RoboNet-1.0 Network of Robotic Telescopes}

RoboNet-1.0 is a project to use a network of three large robotic telescopes: the LT and two Faulkes Telescopes,
which are almost exact clones of the LT. The Faulkes Telescope North (FTN) is situated at Maui in Hawaii
and has been operating since the end of 2003. Faulkes Telescope South (FTS), situated at Siding Spring, Australia,
achieved first light in September 2004. They are financed mainly by 
the Dill Faulkes Educational Trust and are intended for use by UK schools. They are usually in the remote control
mode (operated through the Telescope Management Centre in Liverpool or from schools), but can also operate in 
the fully robotic mode, identical to the LT's. Most of the observing time is
intended for use by UK school children, but some time is available also to the research community. 

This is the core of the 
RoboNet-1.0 project , which is funded by the UK PPARC and includes members of 10 UK university teams in 
Cardiff, Exeter, Hertfordshire, Leicester, Liverpool JMU, Manchester, MSSL, QUB, St. Andrews and 
Southampton. Funds were approved for acquisition of time on FTN and FTS 
for observations on extra-solar
planets and GRBs (in the latter case, 275 hours over next 2.5 years, which covers most of Swift's expected lifetime).
The principal technological aim of the project is to integrate LT, FTN and FTS into a global network of telescopes to act as a single instrument. 
The primary research areas of the project, including GRBs, will greatly benefit from the increased sky and time coverage provided by such a network.

\begin{theacknowledgments}
The Liverpool Telescope is funded via EU, PPARC, JMU grants and the benefaction of Mr. A. E. Robarts.
AG and CG acknowledge the receipt of the Marie Curie Fellowship from the EU and MFB is grateful to the 
UK PPARC for provision of a Senior Fellowship.
\end{theacknowledgments}

\end{document}